\documentstyle[12pt]{article}
\newcommand{\MM}{{\bf M}}

\begin{document}
\baselineskip=0.7cm

\bigskip
\begin{flushleft}
{\Large \bf 
Polarization observables in dp backward elastic scattering
at high and intermediate energies
}

\bigskip
M. Tanifuji$^{a,}$,
S. Ishikawa$^{a,}$\footnote{E-mail: ishikawa@fujimi.hosei.ac.jp} and 
Y. Iseri$^b$

\bigskip
${}^a${\it Department of Physics, Hosei University,
Fujimi 2-17-1, Chiyoda, Tokyo 102, Japan}\\
${}^b${\it Department of Physics, Chiba-Keizai College, 
Todoroki-cho 4-3-30, Inage-ku, Chiba 263, Japan}\\
\end{flushleft}

\bigskip
\bigskip
{\bf Abstract}

The tensor analyzing power $T_{20}$ and the polarization transfer 
coefficients $\kappa_0 (= \frac32 K_y^y)$ and $K_{xz}^y$ are 
investigated for dp backward elastic scattering by the 
invariant-amplitude method. Discrepancies between the conventional 
calculations and the experimental data on $T_{20}$ and $\kappa_0$ 
at high and intermediate energies are mostly dissolved by including 
imaginary parts in the amplitudes. The quantity $K_{xz}^y$ is shown 
to be useful in  criticizing nuclear force assumptions.

{\it PACS:} 21.30.-x, 21.45.+v, 24.70.+s, 25.10.+s

{\it Keywords:} Polarization observable; dp backward elastic scattering; One nucleon exchange model; Invariant amplitude

\newpage
Polarization phenomena in few-body systems are important sources of 
information on nuclear forces and related dynamics.
In particular, the tensor analyzing power $T_{20}$ and the 
polarization transfer coefficient from deuterons to protons $\kappa_0 
(= \frac32 K_y^y)$ in backward elastic scattering of the deuteron by 
the proton at high and intermediate energies  have attracted attention 
because of serious discrepancies between the theoretical 
prediction \cite{Va73,Ko94} 
and the recent experimental data \cite{Pu95}.
For example, the quantities calculated by the PWIA with the one 
 nucleon exchange (ONE) model \cite{Va73,Ko94}, which describes the 
dominant mechanism at backward angles, satisfy the equation of a 
circle in the $\kappa_0-T_{20}$ plane \cite{Ko94}, while the measured 
ones deviate remarkably from the circle along a 
spiral-like curve.
The observables for the inclusive deuteron breakup also suffer 
from similar difficulties \cite{Pu95}.
Several theoretical investigations \cite{Si94,Rek95}, which include QCD 
effects for example, have been attempted but the puzzle still 
remains to be unsolved.

In the present note, we will derive  formulae of the polarization 
observables, $T_{20}$, $\kappa_0$ and $K_{xz}^y$,  for the dp backward 
elastic scattering by the invariant-amplitude method  \cite{Ta68} in 
the non-relativistic framework, assuming the ONE mechanism.
Using the formulae, where the observables are described 
in terms of the invariant amplitudes, we investigate general effects 
of imaginary parts of the  amplitudes, evaluating the magnitudes of 
the amplitudes  by the PWIA. The imaginary parts produce important 
effects on the observables and most of the discrepancies discussed 
above can  be dissolved by the effects.
Recently, model-independent formulae of $T_{20}$ and $\kappa_0$  
in (d,p) reactions have been derived \cite{Ta95} by the method 
similar to the present one.
However, they are applicable to the present scattering only at 
low energies because of additional approximations.
The present work extends the theory to treat the scattering in a 
way free from such approximations. 

The invariant-amplitude method gives  T-matrix ({\MM}) elements 
for a reaction aA$\rightarrow$bB, 
\begin{eqnarray}
\langle\nu_b,\nu_B ; {\bf k}_f | \MM
      | \nu_a,\nu_A; {\bf k}_i\rangle 
  &=& \sum_{s_i s_f K} (s_a s_A \nu_a \nu_A|s_i \nu_i)
 (s_b s_B \nu_b \nu_B|s_f \nu_f)(-)^{s_f - \nu_f}
     (s_i s_f \nu_i -\nu_f|K \kappa) 
\nonumber\\
& & \times \sum_{l_i=\bar{K}-K} ^K [C_{l_i} (\hat{k}_i) \otimes
     C_{l_f=\bar{K}-l_i} (\hat{k}_f)]_{\kappa} ^K F(s_i,s_f,K,l_i),
\label{eq1}
\end{eqnarray} 
where $C_{lm}$ is related to $Y_{lm}$ as usual. The quantity $s(\nu)$ 
is the spin(z-component), $K(\kappa)$ denotes the 
rank (z-component) of tensors which classifies the transition 
amplitudes according to the tensorial character in the spin space,
$F(s_i,s_f,K,l_i)$ is the invariant amplitude which is a 
function of scattering angle $\theta$ and the CM energy, and $\bar{K}$ 
is $K$ for $K=$even and $K+1$ for $K=$odd when 
the parity is unchanged as in the present case. 

The non-vanishing matrix elements at $\theta=\pi$ are the following 
four independent ones which have been derived by the helicity 
conservation in Ref. \cite{Rek95}. At the present, we will calculate 
them by the use of Eq. (\ref{eq1}). In the coordinate system, 
y $\parallel {\bf k}_i \times {\bf k}_f$ and z $\parallel {\bf k}_i$, 
\begin{equation}
\langle 1,\frac12 | \MM | 1,\frac12 \rangle = \frac12 (U_2 + T_2),
\label{eq2}
\end{equation}
\begin{equation}
\langle 1,-\frac12 | \MM | 1,-\frac12 \rangle = \frac{\sqrt{2}}3 U_1 
+\frac16 U_2 -\frac23 T_1 -\frac16 T_2,
\label{eq3}
\end{equation}
\begin{equation}
\langle 1,-\frac12 | \MM | 0,\frac12 \rangle = \langle0,\frac12 
 | \MM | 1,-\frac12\rangle 
= -\frac13 U_1 +\frac1{3\sqrt{2}} U_2 - \frac1{3\sqrt{2}} T_1 
-\frac1{3\sqrt{2}} T_2,
\label{eq4}
\end{equation}
\begin{equation}
\langle 0,\frac12 | \MM | 0,\frac12 \rangle = \frac1{3\sqrt2} U_1 +
 \frac13 U_2 +\frac23 T_1 -\frac13 T_2,
\label{eq5}
\end{equation} 
where ${\bf k}_i$ and ${\bf k}_f$ in the bracket are discarded to 
avoid confusions.  Here $U_j (j=1,2)$ and $T_j (j=1,2)$ are the 
scalar amplitudes and the second-rank tensor ones, respectively, 
and the scalar ones (tensor ones) describe the scattering by the 
spin-space scalar (tensor) interactions. 
The tensor amplitudes include effects of the D-state admixture
in the deuteron ground-state wave function.
They are given as
\begin{equation}
U_j = F ( \frac{2j-1}2,\frac{2j-1}2,0,0 ) ,
\label{eq6}
\end{equation}
\begin{equation}
T_j = F(\frac32,\frac{2j-1}2,2,0) -\sqrt{\frac23} 
F(\frac32,\frac{2j-1}2,2,1) +F(\frac32,\frac{2j-1}2,2,2). 
\label{eq7}
\end{equation}
Here, we will assume the ONE mechanism, for which 
$\langle1,-\frac12 | \MM | 1,-\frac12\rangle$ will vanish because 
the spin-down proton in the incident channel cannot form the spin-up 
deuteron in the final channel due to the lack of the spin flip of 
the proton as discussed below.  
In Eq. (\ref{eq3}), the contribution of the central interactions to 
$U_j$ do not give the spin flip and the contribution
of the second order of the tensor interactions is cancelled
by the residual terms, $-\frac23 T_1 -\frac16 T_2$, in the PWIA limit.  
To take account of this nature of 
$\langle1,-\frac12 | \MM |1,-\frac12\rangle$, we will impose the 
condition,
$\frac{\sqrt{2}}3 U_1 + \frac16 U_2 - \frac23 T_1 - \frac16 T_2 = 0$, 
on the transition amplitudes.  Eliminating $U_1$ by this condition, 
physical quantities are described in terms of  one scalar amplitude 
$U$ and two tensor ones $T$ and $T'$ defined by
\begin{equation}
U = \frac9{2\sqrt2} U_2, \qquad T = -T_1 +2T_2, \qquad 
T^{\prime}=T_1 + {\frac14} T_2. 
\label{eq10}
\end{equation} 

For polarization observables, one can reduce the number of the variables
by introducing the relative magnitudes and phases between $U$, $T$ and 
$T^\prime$,
\begin{equation}
R = \frac{|T|}{|U|},\;R^\prime=\frac{|T^\prime|}{|U|}, \qquad 
\Theta = \theta_T -\theta_U,\;\Theta^\prime = 
\theta_{T^\prime} -\theta_U.
\label{eq15}
\end{equation}
Then we get
\begin{equation}
T_{20} = \{2\sqrt2 R \cos \Theta -R^2 -32{R^\prime}^2 +12R{R^\prime} 
\cos ({\Theta}^\prime - \Theta)\}/N_R,
\label{eq16}
\end{equation}
\begin{equation}
\kappa_0 = \{\sqrt2 -R \cos \Theta -4R^\prime \cos {\Theta}^\prime 
-3\sqrt2 R {R^\prime} \cos ({\Theta}^\prime - \Theta) -
30\sqrt2 {R^\prime}^2\}/N_R 
\label{eq17}
\end{equation} 
with
\begin{equation}
N_R = \sqrt2 +2\sqrt2 R^2 +34\sqrt2 {R^\prime}^2 -4R^\prime 
\cos {\Theta}^\prime. 
\label{eq18}
\end{equation} 
These formulae are exact and independent of details of the reaction 
dynamics except for the restriction by the ONE mechanism. 
The quantities $R$, $R^\prime$, $\Theta$ and $\Theta^\prime$ can 
be treated as free parameters and will 
be determined by experimental data of four independent 
polarization observables.  The parameters thus obtained  will be 
useful for finding or criticizing theoretical models as phase shifts 
 are in the usual scattering \cite{Ta92}.

The experimental data available at the present are not sufficient 
for the determination of the four parameters.  In the following, 
we will calculate $R$ and $R^\prime$ by the PWIA which is 
fundamentally acceptable at high energies, and treat
 $\Theta$ and $\Theta^\prime$ as free parameters 
in the range 
$-180^{\circ} \le \Theta,  \Theta^\prime \le 180^{\circ}$, 
by which imaginary parts are included in the invariant amplitudes. 
The PWIA amplitudes are described by $u(k)$ and $w(k)$, 
the Fourier transforms of the S and D components of the deuteron 
internal wave function. By calculating the LHS of Eqs. (\ref{eq2}), 
(\ref{eq4}), (\ref{eq5}) by the PWIA,
\begin{equation}
U=\frac9{\sqrt2} \{ u^2(k) + \frac14 w^2(k) \} t(k),
\label{eq19} 
\end{equation}
\begin{equation}
T=\frac9{\sqrt2} u(k) w(k) t(k), \qquad
T^ \prime= \frac98 w^2(k) t(k),
\label{eq21}
\end{equation} 
where $t(k)$ is the proton-neutron scattering amplitude at the 
momentum $k$. Denoting $w(k)/u(k)$ by $r$, we get $R$ and 
$R^ \prime$ in the PWIA limit  
\begin{equation}
R = \frac{4|r|}{4 + r^2}
\quad \mbox{and} \quad 
R^\prime = \frac{r^2}{{\sqrt2} (4 + r^2)}.
\label{eq22}
\end{equation} 
As is shown for typical inter-nucleon potentials 
\cite{La80,Re68,Na78,Ma89} in Fig. \ref{fig1}, $r$ decreases from 
zero to minus infinity with the increase of $k$, changes its sign 
at the zero point of $u(k)$, $k=k_0$, and beyond $k_0$  decreases 
from plus infinity. Correspondingly, in the PWIA, 
$\Theta=180^{\circ}$ for $k < {k_0}$ and $\Theta=0^{\circ}$ for 
$k > {k_0}$, and $\Theta^{\prime}$ is zero independently of $k$.  
In general case, at $r=0$, 
\begin{equation}
T_{20} = 0 
\quad \mbox{and} \quad
\kappa_0 = 1,
\label{eq23}
\end{equation} 
which define the point X in the $\kappa_0 - T_{20}$ plane  
in Figs. \ref{fig2}(a)-\ref{fig2}(d) and in the limit 
$r \rightarrow \infty$ 
\begin{equation}
T_{20}= -\frac{4{\sqrt2}}{9- \cos \Theta^ \prime}
\quad \mbox{and} \quad
\kappa_0 = -\frac{7+ \cos \Theta^ \prime}{9- \cos \Theta^ \prime}, 
\label{eq24}
\end{equation} 
which are independent of $\Theta$. In the PWIA limit, 
$T_{20}=-\frac{1}{\sqrt2}$ and $\kappa_0=-1$, which define the 
point Y in the $\kappa_0 - T_{20}$ plane. In the following, the 
polarization observables are calculated for given sets of 
$\Theta$ and $\Theta^ \prime$ by the use of 
Eqs. (\ref{eq16})-(\ref{eq18}) and (\ref{eq22}) by varying $r$ 
from zero to infinity and the calculation is extended by replacing 
$|r|$ by $-|r|$ to the region $k \geq k_0$. 

In Fig. \ref{fig2}(a), the calculated $T_{20}$ and $\kappa_0$ are 
plotted in the $\kappa_0 - T_{20}$ plane for several $\Theta$, 
where $\Theta^ \prime$ is fixed to zero.  The calculated 
quantities are independent on the sign of $\Theta$ due to 
Eqs. (\ref{eq16})-(\ref{eq18}). For $\Theta^ \prime=0^{\circ}$, 
the present Eqs. (\ref{eq16}) and (\ref{eq17}) are reduced to 
Eqs. (16) and (17) in Ref. \cite{Ta95}, respectively, when $r$ 
is replaced again by $R$ after the transformation by Eq. (\ref{eq22}). 
Then, Fig. \ref{fig2}(a) is essentially the same as Fig. 3 in the 
reference. As was discussed there, the point P defined by a set 
of $\kappa_0$ and $T_{20}$ calculated by the PWIA for an arbitrary 
$k$ moves clockwise along the circle denoted by 
$\Theta=180^ \circ$, from X to a certain point through Y  with 
the increase of $k$.  The trajectories of the point P similarly 
defined for other $\Theta$ are deformed toward the inside of the 
circle according to the decrease of $\Theta$ from $180^{\circ}$ to 
$90^{\circ}$, where the point for $k=k_0$ is fixed to Y.  We call 
this deformation of the trajectories the "$\Theta$ effect".  The 
experimental data \cite{Pu95} for the small $k$ are mostly located 
between the two lines for $\Theta=120^ \circ$ and $135^ \circ$.  
Then the $\Theta$ effect is important to describe such small 
$k$ data. 

Figs. \ref{fig2}(b) and \ref{fig2}(c) show effects of finite 
$\Theta^ \prime$ for $\Theta^\prime > 0^{\circ}$ and for 
$\Theta^\prime < 0^{\circ}$, respectively, where $\Theta$ is 
fixed to $135^ \circ$. In Fig. \ref{fig2}(b), the calculations for 
$\Theta^ \prime$ larger than $90^\circ$ are ignored to avoid the 
complication of the figure. In both of Figs. \ref{fig2}(b) and 
\ref{fig2}(c), the point P($k_0$) which is defined by 
$\kappa_0$ and $T_{20}$ given by Eq. (\ref{eq24}) moves on the 
X--Y straight line from Y toward the center of the circle with the 
increase of the magnitude of $\Theta^ \prime$,  accompanied by 
the corresponding deformation of the trajectories. 
We call this the "$\Theta^ \prime$ effect". 
To reproduce the data for the large $k$, the $\Theta^ \prime$ effect 
is clearly important. In the cases of $\Theta^ \prime=30^\circ$  
and $-105^\circ$, for example, the agreement between the calculation 
and the experiment is much improved compared with the case of the 
PWIA. To see the pure $\Theta^ \prime$ effect, the calculations for 
several $\Theta^ \prime$ with $\Theta=180^{\circ}$ are performed, 
the results of which are shown in Fig. \ref{fig2}(d), where the 
calculated are independent on the sign of $\Theta^ \prime$. The 
$\Theta^ \prime$ effect is quite remarkable for large magnitudes 
of $\Theta^ \prime$. Most of the data for the large $k$ are 
located between the trajectories calculated with 
$\Theta^\prime=105^\circ$ and $120^ \circ$, although they miss the 
agreements with the small $k$ data . For further investigations, 
we will classify the trajectories of P in these figures, according 
to their gross behaviour, into the "egg shape", the "cusp" upon 
the X--Y line and the "{\it 8} shape".  
In Fig. \ref{fig2}(b), for example, the trajectory for 
$\Theta^ \prime=30^\circ$ 
is the egg shape, those for $\Theta^\prime=60^\circ$ and $90^\circ$ 
are the cusp and the {\it 8} shape. 

To examine the $\Theta$ and $\Theta^ \prime$ effects in more detail, 
we will plot $T_{20}$ and $\kappa_0$ calculated by 
Eqs. (\ref{eq16})-(\ref{eq18}) and (\ref{eq22}) with 
the Paris potential \cite{La80} as the function of $k$. 
The shown in Figs. \ref{fig3}(a) and \ref{fig3}(b) are for the 
typical trajectories in Figs. \ref{fig2}(b)-\ref{fig2}(d), i.e. 
for the sets $(\Theta,\Theta^ \prime)=$ ($135^\circ$, $30^\circ$), 
($135^\circ$, $-105^\circ$), ($180^\circ$, $105^\circ$), 
($180^\circ$, $120^\circ$)  and the pure PWIA. The trajectories for 
the first two sets are the egg shape and those for the third and 
the fourth are the cusp and the {\it 8} shape, respectively. 
In Fig. \ref{fig3}(a), the calculations for the former two sets 
do not give the structure of $T_{20}$, 
which is observed experimentally as a local maximum in the range 
$k=0.25-0.45$ GeV/c, 
while those for the latter two produce the structures similar to the 
experimental one. To produce the structure, even in 
Figs. \ref{fig2}(b)-\ref{fig2}(d) $T_{20}$ is required to have a 
maximum in the range $\infty > r > 0$. In 
Figs. \ref{fig2}(b)-\ref{fig2}(d), such a maximum of $T_{20}$ is seen 
in the third quadrant of the $\kappa_0$--$T_{20}$ plane 
for  the trajectories of the cusp type and the {\it 8} shape type, 
though unclear because of the broad shape, while the maximum is not 
seen for those of the egg shape. 
The experimental data behave like a cusp upon the X--Y line although 
modified by the two factors, the fluctuation of 
$\kappa_0$ with $k$ in the large $k$ region and the small bump of 
$T_{20}$ around $k=0.44$ GeV/c. In Fig. \ref{fig2}(d), such 
features of the data are demonstrated by connecting the data points 
by straight lines.  In Fig. \ref{fig3}(a), the calculations by the 
sets ($135^\circ$, $30^\circ$) and ($135^ \circ$, $-105^\circ$) 
describe the $T_{20}$ data very well except the structure. 
In Fig. \ref{fig3}(b), the calculated $\kappa_0$ is compared with 
the experimental data, where the present calculations except 
the one for the set ($135^\circ$, $30^\circ$) give agreements 
with the data better than those in the PWIA, for the large $k$.  
Similar numerical calculations are performed for 
the RSC \cite{Re68}, Nijmegen \cite{Na78} and Bonn B \cite{Ma89} 
potentials. As is speculated from their features seen in 
Fig. \ref{fig1}, 
the distributions of $T_{20}$ and $\kappa_0$ versus $k$ obtained by 
the Bonn B potential are considerably stretched toward the larger $k$,
while those calculated by other potentials are rather similar to 
those by the Paris potential. 

In the present investigation, we assume $\Theta$ and $\Theta^ \prime$ 
to be independent of $k$ for the convenience of examining the general
 effect of the imaginary part of the amplitudes, although there is 
no justification for such assumptions. To reproduce the experimental 
data more quantitatively, one will vary $\Theta$ and $\Theta^\prime$
 with $k$. For example, vary $\Theta$ from $135^\circ$ to $180^\circ$  
with the increase of $k$ and choose $\Theta^\prime$ between 
$-105^\circ$ and $-120^\circ$. In such cases, experimental data of 
other observables will be necessary to solve possible ambiguities of 
the parameters. A candidate of such observables is the deuteron-proton 
polarization transfer coefficient $K_{xz}^y$, 
\begin{equation}
K_{xz}^y= 3\{-R \sin \Theta + 5\sqrt2 R{R^ \prime} 
\sin (\Theta - \Theta^ \prime)\}/{N_R}.
\label{eq25}
\end{equation} 
The quantity vanishes in the PWIA limit and is sensitive to 
$\Theta$ and $\Theta^ \prime$ in general case. For example, its 
sign is changed by the change of the sign of $\Theta^ \prime$ when 
$\Theta=180^{\circ}$. Fig. \ref{fig3}(c) shows  $K_{xz}^y$ 
calculated for the ($\Theta, \Theta^ \prime$) sets same as those in 
Figs. \ref{fig3}(a) and \ref{fig3}(b). 
The quantity calculated for the set ($135^\circ, 30^\circ$) 
and that for the set ($135^\circ, -105^\circ$) behave quite 
differently from each other with the opposite sign except those 
at the small $k$, contrary to their similarity in $T_{20}$. 
 Also $K_{xz}^y$ will be useful 
in criticizing the nuclear force assumptions, because it vanishes 
at $k=k_0$ and $k_0$ depends on the force assumption, for example 
$k_0=0.39$ and $0.45$ GeV/c for the RSC potential and the Bonn B 
one, respectively. 

From the features of the $\Theta$ and $\Theta^{\prime}$ effects,
one can speculate about the dynamical origin of the effects. 
Since the $\Theta$ effect is important in the small $k$ region, 
i.e. at low incident energies, it will originate from 
non-mesonic phenomena like virtual breakup of the deuteron, 
the effect of which becomes less important at high energies in 
the usual deuteron scattering \cite{Ya86}. At energies higher 
than the pion 
threshold, mesonic effects, which include excitations of $\Delta$ 
and other baryon resonances, will become important. These will 
be responsible for the $\Theta^ \prime$ effect. Earlier mesonic 
contributions have been investigated in Refs. \cite{Bo85,Na85}, 
where the calculations produce some structures of $T_{20}$ which 
correspond to the one discussed above but the calculated $T_{20}$ 
at the minimum, which appears around $k \simeq 0.3$ GeV/c, 
is too much negative compared 
with the new data \cite{Pu95}, which are considerably less 
negative than the old ones  \cite{Ar84} in that region of $k$. 
The latter feature of the calculations may be related to the 
insufficient treatment of the 
$\Theta$ effect.  Finally, the application to 
the inclusive deuteron breakup and the examination of relativistic 
effects are now in progress.

\bigskip
\bigskip
The authors wish to thank Professors Y. Koike amd T. Hasegawa for 
valuable discussions and are indebted to Professor C. F. Perdrisat 
for providing the details of his data. 

\newpage

\newpage

\begin{figure}
\caption{
Fourier transforms of deuteron internal wave functions. 
$u(k)$ and $w(k)$ are for the S and D components and $k$ is 
the p-n relative momentum. The calculated are for the 
RSC (dash-dotted lines), Nijmegen (dotted lines), 
Paris (solid lines), and Bonn B (dashed lines) potentials. 
The zero point of $u(k)$, $k=k_0$, is shown by the arrow  
for the Paris potential for example.
\label{fig1}}
\end{figure}

\begin{figure}
\caption{
$T_{20}$ versus $\kappa_0$ (=$\frac32 K_y^y$). 
The experimental data are for backward elastic scattering 
(open circles) and inclusive breakup (solid circles, only in (a) ) 
\protect\cite{Pu95}. The curves are calculated by 
Eqs. (\protect\ref{eq16})-(\protect\ref{eq18}) and 
(\protect\ref{eq22}) for $\Theta$ = 180$^{\circ}$, 135$^{\circ}$, 
120$^{\circ}$, 90$^{\circ}$ with $\Theta ^{\prime} = 0^{\circ}$ 
in (a), for $\Theta^{\prime}$ = 30$^{\circ}$, 60$^{\circ}$, 
90$^{\circ}$ with $\Theta$=135$^{\circ}$ in (b), for 
$\Theta^{\prime}$ = $-60^{\circ}$, $-105^{\circ}$, $-150^{\circ}$ 
with $\Theta$ = 135$^{\circ}$ in (c) and for 
$\Theta^{\prime}$=60$^{\circ}$, 105$^{\circ}$, 120$^{\circ}$ with 
$\Theta$ = 180$^{\circ}$ in (d). The large circles are the PWIA 
calculation. In (d), the data points are connected by straight 
lines (see text).  
\label{fig2}}
\end{figure}

\begin{figure}
\caption{
$T_{20}$, $\kappa_0$ and $K_{xz}^y$ versus $k$. 
The open circles are the experimental data for the backward 
elastic scattering \protect\cite{Pu95}. The curves are calculated by 
Eqs. (\protect\ref{eq16})-(\protect\ref{eq18}) and 
(\protect\ref{eq22}) with the Paris potential, where the lines are 
for ($\Theta$, $\Theta^{\prime}$)=(135$^{\circ}$, 30$^{\circ}$) 
[the solid], (135$^{\circ}$, $-105^{\circ}$) [the dashed], 
(180$^{\circ}$, 105$^{\circ}$) [the dotted] and 
(180$^{\circ}$, 120$^{\circ}$) [the dash-dotted]. The thin solid 
lines are the PWIA calculation, which gives zero for $K_{xz}^y$. 
The vertical dash-dotted straight line indicates the location of 
$k=k_0$.
\label{fig3}}
\end{figure}

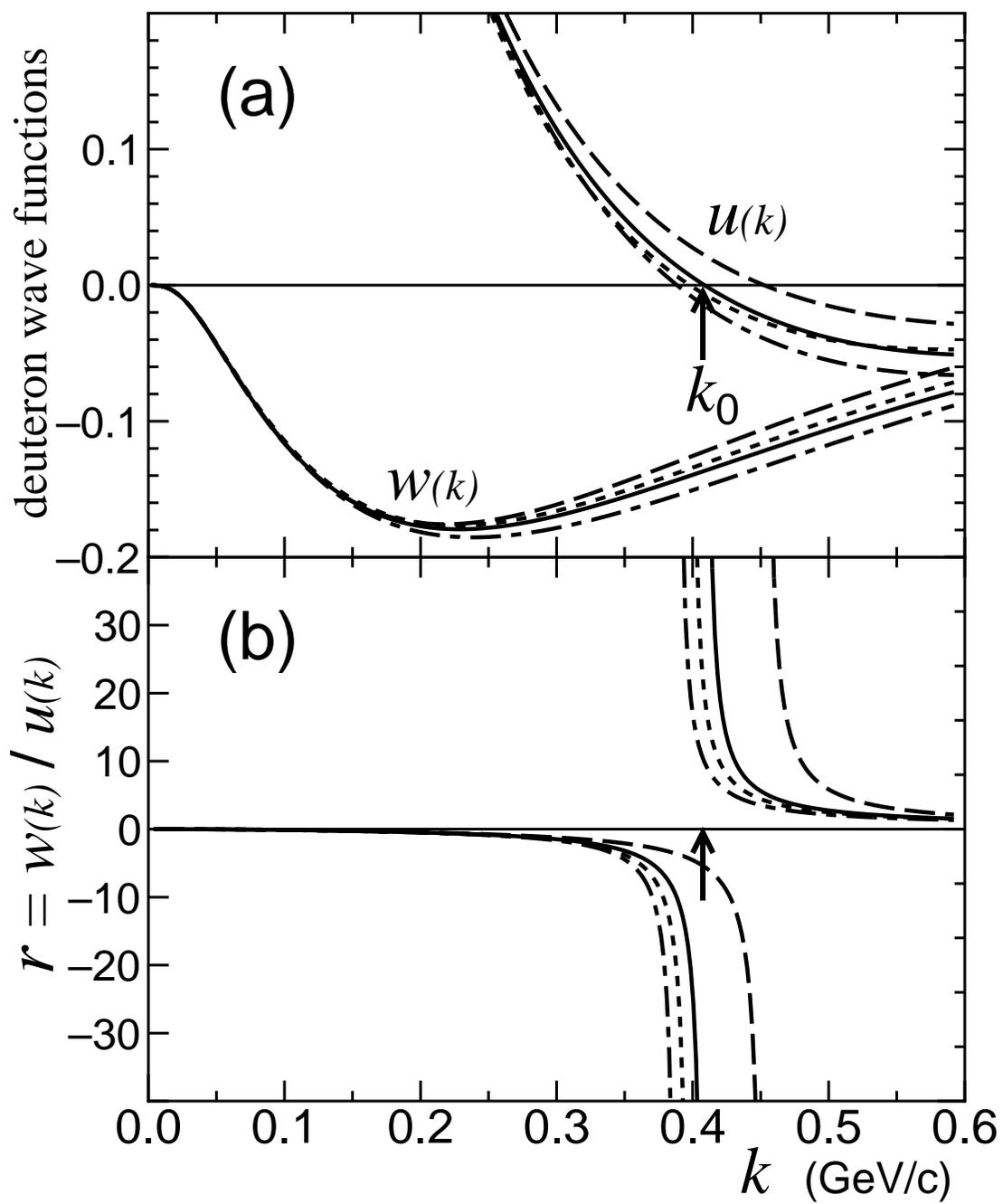
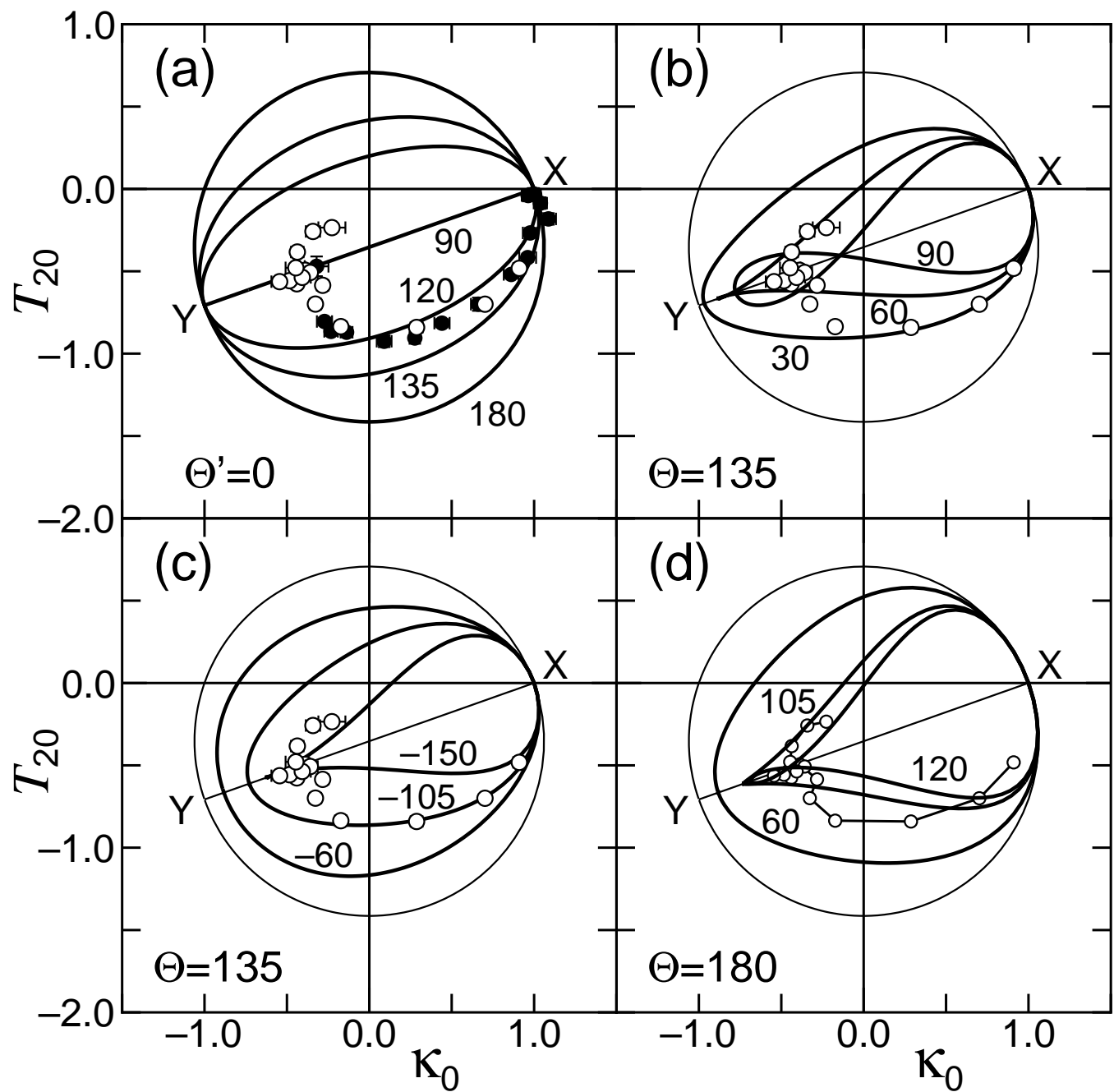
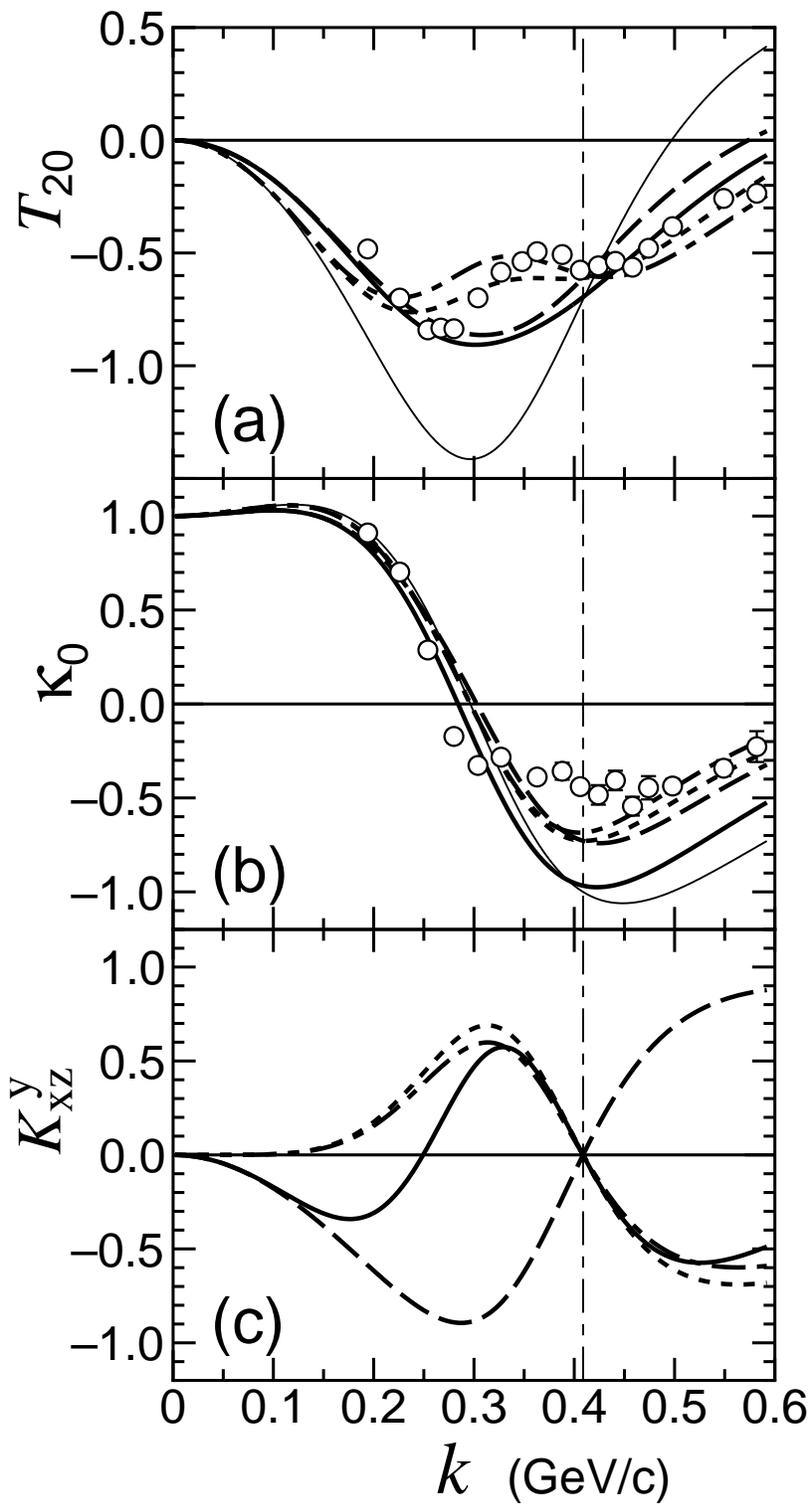
\end{document}